\documentclass[aps,showpacs,twocolumn]{revtex4}
\usepackage{graphicx}
\usepackage{latexsym}
\usepackage{bm}
\begin{document}
\newcommand{\jpb}{{J. Phys. B: At. Mol. Opt. Phys. }} 
\title{Equivalence of  the velocity and length gauge perturbation series}
\author{F.H.M. Faisal}
\affiliation{Fakult\"at f\"ur Physik, Universit\"at Bielefeld, Postfach 100131, 
D-33501 Bielefeld, Germany}
\begin{abstract} 
We derive a ``master'' perturbation expansion for the quantum transition amplitude in a light field between the field-free initial and final atomic states in the minimal-coupling (MC) ``velocity'' gauge. The result is used to prove that the traditional ``velocity'' and ``length'' gauge perturbation series are equivalent infinite series representations or branches of the same amplitude function, that are equal but in a common domain of convergence (if it exists). More generally, we show that they constitute only two members of a one-parameter family of infinitely many branches of the given transition amplitude.
\end{abstract}


\maketitle

A long-standing unresolved problem of the quantum mechanical perturbation theory of light-matter interaction is the existence of two distinct infinite perturbation series for a given transition amplitude in the so-called ``velocity'' and ``length'' gauges, that have hitherto resisted a demonstration of their mathematical equivalence. 
In addition, numerical calculations of the transition probability based on the two series (albeit, for practical reasons, only in their truncated forms) have frequently shown a significant discrepancy between them and/or with various experimental data. These and related difficulties 
have led some authors to argue in favor of the length gauge (e.g. \cite{Lamb, Yang}) as opposed to the velocity gauge, since the former is based manifestly on a  physically ``true'' energy operator \cite{Cohen-Tannoudji}.
Nevertheless, the principle of gauge invariance in {\it quantum} theory (e.g. \cite{Fock, Cohen-Tannoudji, Faisal}) requires that they aught to be equivalent.

The purpose of this Letter is to derive a ``master'' perturbation expansion of the quantum mechanical
transition amplitude in the minimal-coupling (MC) ``velocity'' gauge, 
and to use the result to demonstrate that the two perturbation series, that are 
traditionally obtained in the ``velocity'' and ``length'' gauges, are two branches of the same amplitude function and, hence, they are equal but in a common domain of convergence (if there is any).
More generally, our result shows that they belong to a one-parameter family of 
infinitely many equivalent series representations (or branches) of the
transition amplitude for a given transition process.
      
The Schr\"odinger equation of an atomic system interacting with an electromagnetic field,
in the minimal-coupling (MC) transverse gauge
is given by: 
\begin{equation}\label{TotSchEqu}
(i\hbar \frac{\partial}{\partial t} - H_{MC}(t))\Psi_{{MC}}(t) =0
\end{equation}
where, the total Hamiltonian of the interacting system is
\begin{equation}\label{MinCouHam}
H_{MC}(t) = \frac{(\bm{p}_{op} - \frac{e}{c} \bm{A}(t))^{2}}{2m} + e A_{0}(\bm{r})\end{equation}
with the four-potential 
$A_{\mu} \equiv (A_{0}(\bm{r}), \bm{A}(t))$
where, the scalar potential can be used to define the  ``atomic'' potential, 
$eA_0(\bm{r}) = V_{a}(\bm{r})$, and
$ \bm{A}(t)$ is the transverse vector potential of the light field.
In the usual electric dipole approximation
(e.g. Bohr-radius/ wavelength $ << 1$), 
the vector potential $\bm{A}(t)$ depends only on $t$.
As usual, the ``atomic'' Hamiltonian
\begin{equation}\label{AtoHam}
H_{a} \equiv \frac{\bm{p}_{op}^{2}}{2m} + V_{a}(\bm{r})
\end{equation}
provides a complete set of eigen-states
\begin{equation}\label{AtoHam}
\sum_{all j}|\phi_{j}^{a}><\phi_{j}^{a}| = 1 
\end{equation}
and eigen-energies, $E^{a}_{j}$,
which satisfy the eigenvalue equation:
\begin{equation}
H_{a}|\phi_{j}^{a}> = E_{j}|\phi_{j}^{a}>, \text{all j}.
\end{equation} 
We define the ``atomic'' Green's function $G_{a}(t,t')$ by  the equation 
\begin{equation}\label{AtoGreEqu}
(i\hbar\frac{\partial}{\partial t} - H_{a}(t))G_{a}(t,t') = \delta(t-t').
\end{equation} 
Its solution is given by,
\begin{eqnarray}\label{AtoGreFun}
G_{a}(t,t') &= &-\frac{i}{\hbar}\theta(t-t')e^{-\frac{i}{\hbar}H_{a}(t-t')}\nonumber\\
&=& -\frac{i}{\hbar}\theta(t-t')\sum_{all j}|\phi_{j}^{a}>e^{-\frac{i}{\hbar}E_{j}(t-t')}<\phi_{j}^{a}|\nonumber\\
\end{eqnarray}
This can be easily verified by its substitution in Eq. (\ref{AtoGreEqu}) and noting that the derivative of the theta-function is the delta function. 
Finally, we define, for later use, the total Green's function (or propagator) $G_
{MC}(t,t')$ associated with the minimal-coupling Hamiltonian $H_{MC}(t)$, by the inhomogeneous differential equation:
\begin{equation}\label{TotGreEqu}
(i\hbar\frac{\partial}{\partial t} - H_{MC}(t)) G_{MC}(t,t')= \delta(t-t').
\end{equation}

The total Hamiltonian $H_{MC}(t)$ can always be written
as a sum of two terms, in infinitely many ways:
\begin{eqnarray}\label{SumHam}
H_{MC}(t) &=& H_{s}(t) + V_{s}(t) ;  s=1,2,3,\cdots \infty
\end{eqnarray}
where, $V_{s}(t) \equiv H_{MC}(t) - H_{s}(t)$, and the
basis Hamiltonians $H_{s}(t)$ can be used to define the associated basis Green's functions, $G_{s}(t,t')$, 
by the equations:
\begin{equation}\label{BasGreEqu}
(i\hbar \frac{\partial}{\partial t} - H_{s}(t))G_{s}(t,t') =\delta(t-t'); s=1,2,3 \cdots \infty.
\end{equation}
Eq. (\ref{BasGreEqu}) can be solved in terms of the linearly independent  complete set of fundamental solutions, $|\phi_{j}^{(s)}(t)>$, for each $s$ and all $j$,  of the homogeneous Schr\"odinger equation,
 \begin{equation}\label{BasSchEqu}
(i\hbar\frac{\partial}{\partial t} - H_{s}(t)) |\phi_{j}^{(s)}(t)> =0; s=1,2,3, \cdots, \infty; all j. 
\end{equation}
Explicitly, we have:
\begin{equation}\label{BasGreFun}
G_{s}(t,t') =-\frac{i}{\hbar}\theta(t-t')\sum_{all j} |\phi_{j}^{(s)} (t)><\phi_{j}^{(s)} (t')|.  
\end{equation}
We may now expand the total Green's function, $G_{MC}(t,t')$, using any basis Green' function, $G_{s}(t,t')$, in an infinite series,
\begin{eqnarray}\label{TotGreFun}
G_{MC}(t,t') &=& G_{s}(t,t') +
\int dt_1G_{s}(t,t_1)D_{MC}(t_1,t')\nonumber\\
&\times&G_{s}(t_1,t') + \int \int dt_{1}dt_2 G_{s}(t,t_1)\nonumber\\
&\times&D_{MC}(t_1,t_{2})
G_{s}(t_1,t_2)D_{MC}(t_2,t')G_{s}(t_2,t')\nonumber\\
&+&  \cdots; s=1,2,3,\cdots,\infty. 
\end{eqnarray}
where we have introduced the (inhomogeneous Schr\"odinger) operator: 
\begin{equation}\label{SchOpe}
D_{MC}(t,t')\equiv [(H_{MC}(t) - i\hbar\frac{\partial}{\partial t}) + i\hbar\delta(t-t')]
\end{equation} 
which, it is worth noting, depends on the {\it total} Hamiltonian $H_{MC}(t)$, and not on its various partitions. 
The symbol $\int$
above stands for the integration over the entire time axis:
$\int \equiv \int_{-\infty}^{\infty}$.  We may note, parenthetically, that the presence of the theta-function in the explicit representation of a Green's function, Eq. (\ref{BasGreFun}), automatically accounts for the appropriate domains of the intermediate time integrations in Eq. (\ref{TotGreFun}).
Eq. (\ref{TotGreFun}) is a solution of the Green's equation
(\ref{TotGreEqu}). This can be easily verified as follows:
First,  we apply (\ref{SchOpe}) on (\ref{BasGreFun}) and use (\ref{SumHam}) to get,
\begin{eqnarray}\label{SchOpeIde}
&&D_{MC}(t,t')G_{s}(t,t') \nonumber\\
&=& [(H_{MC}(t) - i\hbar\frac{\partial}{\partial t}) + i\hbar\delta(t-t')]G_{s}(t,t')\nonumber\\   
&=& [(H_{s}(t)-i\hbar\frac{\partial}{\partial t})G_{s}(t,t') + V_{s}(t)G_{s}(t,t') \nonumber\\
&+& i\hbar\delta(t-t')G_{s}(t,t')]\nonumber\\
&=& [-\delta(t-t') +V_{s}(t)G_{s}(t,t') + \delta(t-t')\theta(t-t')\nonumber\\
&\times& \sum_{all j} |\phi_{j}^{(s)}(t)><\phi_{j}^{(s)}(t')| ]\nonumber\\
&=& [-\delta(t-t') +V_{s}(t)G_{(s)}(t,t') +\delta(t-t')\theta(t-t')]\nonumber\\
&=& V_{s}(t)G_{s}(t,t'); s=1,2,3 \cdots,\infty.
\end{eqnarray}
where, $V_{s}(t)\equiv H_{MC}(t)-H_{s}(t)$, and 
in the last step we have used the fact that an integration variable, say $t_{n+1}$, in any of the integrations in the 
series (\ref{TotGreFun}), always approaches its upper limit, say $t_{n}$, as: $\lim (t_{n}-t_{n+1})=0^{+}$, $\theta(t_{n}-t_{n+1}) =1$.
Second, substituting Eq. (\ref{SchOpeIde}) in Eq. (\ref{TotGreFun}) we may sum the series on the right hand side as,
\begin{eqnarray}\label{SerSum}
& & G_{MC}(t,t')\nonumber\\
&=& G_{s}(t,t') + \int dt_{1}G_{s} (t,t_{1})V_{s}(t_{1})\times [G_{s}(t_{1},t')\nonumber\\
&+&\int dt_{2}G_{s}(t_{1},t_{2})V_{s}(t_{2})G_{s}(t_{1},t') +\cdots]\nonumber\\
&=&  G_{s}(t,t') + \int dt_{1}G_{s} (t,t_{1}V_{s}(t_{1})\times [G_{MC}(t_{1},t')] \nonumber\\
\end{eqnarray}
where, we have used the fact that the quantity in the square brackets above is equal to the series itself.
Next, operating on the last equation from the left with $(i\hbar\frac{\partial}{\partial t} - H_{s}(t))$, noting Eq. (\ref{BasGreEqu}), and carrying out the resulting delta-function integration at once, we get,
\begin{equation}
(i\hbar\frac{\partial}{\partial t} - H_{s}(t))G_{MC}(t,t') = \delta(t-t') + V_{s}(t)
G_{MC}(t,t') 
\end{equation}
Finally, on transposing the last term to the left hand side and using Eq. (\ref{SumHam}) we arrive at, 
\begin{equation}
(i\hbar\frac{\partial}{\partial t} - H_{MC}(t))G_{MC}(t,t_{i}) = \delta(t-t')
\end{equation}
which agrees with Eq. (\ref{TotGreEqu}); q.e.d.
Thus, Eq. (\ref{TotGreFun}) provides a general series solution for the total Green's function $G_{MC}(t,t')$ for an {\it arbitrary} 
choice of the basis Hamiltonian $H_{s}(t)$, and the associated basis Green's function $G_{s}(t,t')$. 
We may choose, 
\begin{equation}\label{ParHam}
H_{s}(t) \equiv H_{\lambda}(t)
= \frac{(\bm{p}_{op} -\frac{e}{c}\lambda\bm{A}(t))^{2}}{2m} + V_{a}(\bm{r}) -\lambda\frac{e}{c}\dot{\bm{A}}(t)\cdot\bm{r}
\end{equation}
where $\lambda$ is a real number. The 
associated Green's function 
$G_{\lambda}(t,t')$ is defined by
\begin{equation}\label{ParGreEqu}
(i\hbar\frac{\partial}{\partial t} - H_{\lambda}(t))G_{\lambda}(t,t') = \delta(t-t') 
\end{equation} 
We find its exact solution to be,
\begin{eqnarray}\label{ParGreFun}
G_{\lambda}(t,t')&=&-\frac{i}{\hbar}\theta(t-t')e^{i\lambda \frac{e}{\hbar c}\bm{A}(t)\cdot\bm{r}}\nonumber\\
&\times&\sum_{all j} |\phi_{j}^{a}>
e^{-\frac{i}{\hbar}H_{a}(t-t')} <\phi_{j}^{a}|\nonumber\\
&\times&e^{-i\lambda \frac{e}{\hbar c}\bm{A}(t')\cdot\bm{r'}}\nonumber\\
&=&e^{i\lambda \frac{e}{\hbar c}\bm{A}(t)\cdot\bm{r}}G_{a}(t,t')e^{-i\lambda \frac{e}{\hbar c}\bm{A}(t')\cdot\bm{r'}}
\end{eqnarray}
The validity of the above solution can be established without difficulty by substituting Eq. (\ref{ParGreFun}) in Eq. (\ref{ParGreEqu}), 
and simplifying by noting the definition (\ref{AtoGreFun}) and using,
\begin{equation}
(\bm{p}_{op} -\lambda\frac{e}{c}\bm{A}(t))^{2}e^{i\lambda\frac{e}{\hbar c}\bm{A}(t)\cdot\bm{r}} = e^{i\lambda\frac{e}{\hbar c}\bm{A}(t)\cdot\bm{r}} \bm{p}_{op}^{2}.
\end{equation}

We now substitute $G_{s} (t,t')\equiv G_{\lambda}(t,t')$ in Eq. (\ref{TotGreFun}) for the total Green's function, to get: 
\begin{eqnarray}\label{TotGreFunPri}
& &G_{MC}(t,t')\nonumber\\
&=& (G_{\lambda}(t,t') + \int dt_1G_{\lambda}(t,t_1)D_{MC}(t_1,t')G_{\lambda}(t_1,t')\nonumber\\
&+& \int \int dt_2 dt_1 G_{\lambda}(t,t_1)D_{MC}(t_1,t_{2})G_{\lambda}(t_1,t_2)\nonumber\\
&\times&D_{MC}(t_{2},t')G_{\lambda}(t_2,t') +  \cdots)\nonumber\\
&=& e^{i\lambda \frac{e}{\hbar c}{\bm{A}}(t)\cdot\bm{r}} [ G_{a}(t,t')
+ \int dt_1G_{a}(t,t_1)D_{MC}^{(\lambda)}(t_1,t')\nonumber\\
&\times&G_{a}(t_1,t')+ \int \int dt_1 dt_2 G_{a}(t,t_1)D_{MC}^{(\lambda)}(t_1,t_{2})\nonumber\\
&\times&G_{a}(t_1,t_2)D_{MC}^{(\lambda)}(t_2,t')G_{a}(t_2,t')
+ \cdots ]e^{-i\lambda \frac{e}{\hbar c}\bm{A}(t')\cdot\bm{r'}}\nonumber\\
\end{eqnarray}
where,
\begin{eqnarray}\label{TraSchOpe}
& &D_{MC}^{\lambda}(t,t')\nonumber\\
&\equiv& e^{-i\lambda \frac{e}{\hbar c}\bm{A}(t)\cdot\bm{r}} 
D_{MC}(t,t') e^{i\lambda \frac{e}{\hbar c}\bm{A}(t)\cdot\bm{r}}\nonumber\\
&=&\{ [\frac{(\bm{p}_{op} +(\lambda-1)\frac{e}{c}\bm{A}(t))^{2}}{2m} +V_{a}(\bm{r}) + \lambda\frac{e}{c}\dot{\bm{A}}(t)\cdot\bm{r}\nonumber\\
&-& i{\hbar}\frac{\partial}{\partial t}] + i\hbar\delta(t-t')\}\nonumber\\
&=&\{ [(\frac{\bm{p}_{op}^{2}}{2m}+V_{a}(\bm{r})) - i{\hbar}\frac{\partial}{\partial t}]+V_{\lambda}(t) + i\hbar\delta(t-t')\} \nonumber\\
\end{eqnarray} 
and, 
\begin{eqnarray}\label{ParIntHam}
V_{\lambda}(t)&=& [(\lambda-1)\frac{e}{m c} \bm{A}(t)\cdot\bm{p}_{op} 
+(\lambda-1)^{2}\frac{e^{2}}{2mc^{2}}A^{2}(t)  \nonumber\\
&+& \lambda\frac{e}{c}\dot{\bm{A}}(t)\cdot\bm{r}]
\end{eqnarray}
Operating with $D_{MC}^{\lambda}(t,t')$ from the left on to $G_{a}(t,t')$, noting Eq. (\ref{AtoGreEqu}), and calculating similarly as in the case of Eq. (\ref{SchOpeIde}) above, we get: 
\begin{eqnarray}\label{TraIntHam}
& &D_{MC}^{\lambda}(t,t') G_{a}(t,t')\nonumber\\
&=& \{ [(\frac{\bm{p}_{op}^{2}}{2m} + V_{a}(\bm{r})) -i\hbar\frac{\partial}{\partial t}] G_{a}(t,t')\nonumber\\
&+& V_{\lambda}(t)G_{a}(t,t') +i\hbar\delta(t-t')G_{a}(t,t')\} \nonumber\\
&=& V_{\lambda} (t) G_{a}(t,t'). 
\end{eqnarray}
Finally, using Eq. (\ref{TraIntHam}) in Eq. (\ref{TotGreFunPri}), we obtain
\begin{eqnarray}\label{TotGreFunFin} 
G_{MC}(t,t') &=& e^{i\lambda\frac{e}{\hbar c}\bm{A}(t)\cdot\bm{r}} [ G_{a}(t, t') + 
\int d t_{1}G_{a}(t,t_{1})V_{\lambda}(t_{1})\nonumber\\
&\times&G_{a}(t_{1}, t') 
+\int  \int d t_{1} d t_{2} G_{a}(t,t_{1})V_{\lambda}(t_{1})\nonumber\\
&\times&G_{a}(t_{1},t_{2})V_{\lambda}(t_{2})G_{a}(t_{2},t')\nonumber\\
&+& \cdots] e^{-i\lambda\frac{e}{\hbar c}\bm{A}(t')\cdot\bm{r'}}
\end{eqnarray}
This is a one-parameter family of infinite series representations of the total Green's function $G_{MC}(t,t')$, for any value of the real parameter $\lambda$ : 

The transition amplitude, 
 $S_{MC}^{f\leftarrow i}(t_{f},t_{i})$,
 between the field-free 
 reference states \cite{foot-01}:
  \begin{eqnarray}
 |\phi^{\lambda}_{i}(t_{i})>&=&e^{i\lambda\frac{e}{\hbar c}\bm{A}(t_{i})\cdot\bm{r'}}|\phi^{a}_{i}(t_{i})>\label{IniSta} \\ 
 &\text{and}&\nonumber\\
<\phi^{\lambda}_{f}(t_{f})|&=&<\phi^{a}_{f}(t_{f})|
e^{-i\lambda\frac{e}{\hbar c}\bm{A}(t_{f})\cdot\bm{r}}
\label{FinSta}
\end{eqnarray}
that are prepared initially  
at $t'=t_{i}$  and
detected finally at a later time $t=t_{f}$, where $\bm{A}(t_{i,f})$ are {\it arbitrary} constant vector potentials, 
is:
 \begin{eqnarray}\label{TraAmp} 
& &S_{MC}^{f\leftarrow i} (t_{f},t_{i})\nonumber\\
&\equiv& i\hbar <\phi^{\lambda}_{f}(t_{f})|G_{MC}(t_{f}, t_{i})|\phi^{\lambda}_{i}(t_{i})>\nonumber\\
&=& \delta_{fi}  -\frac{i}{\hbar} [
\int dt_{1} <\phi^{a}_{f}(t_{1})|V_{\lambda}(t_{1})|\phi_{i}^{a}(t_{1})>\nonumber\\
&+& \int \int dt_{1} dt_{2}<\phi^{a}_{f}(t_{1})|V_{\lambda}(t_{1}) G_{a}(t_{1},t_{2})|\phi_{i}^{a}(t_{2})>\nonumber\\
&+& \cdots]; \qquad  (\lambda,\, \text{a real number})
\end{eqnarray}
where, we have used the total Green's function $G(t_{f},t_{i})$ given by 
Eq. (\ref{TotGreFunFin}), cancelled the phase factors depending on the arbitrary (constant) vector potentials, and simplified by putting,
\begin{eqnarray}
<\phi_{f}^{a}(t_{f})|G_{a}(t_{f}, t_{1}) &=& -\frac{i}{\hbar}<\phi_{f}^{a}(t_{1})|\\
&\text{and}&\nonumber\\
G_{a}(t_{n}, t_{i})|\phi_{i}^{a}(t_{i})> &=& -\frac{i}{\hbar}|\phi_{i}^{a}(t_{n})>.
\end{eqnarray}
Note that the above expression (\ref{TraAmp}) holds for any value of the real parameter 
$\lambda$, and thus constitutes a  ``master'' expansion 
of the transition amplitude derived 
in the minimal-coupling (MC)  ``velocity'' gauge.
It provides a one-parameter family of 
infinitely many series representations or branches of
the amplitude function $S_{MC}^{f\leftarrow i}(t_{f},t_{i})$.
We may choose, for instance, the parameter $\lambda = 0, 1, $ or $\frac{1}{2}$ on the right hand side of Eq. (\ref{TraAmp}), and get, respectively, the three expansions of the transition amplitude:
 \begin{eqnarray}
&&S_{MC}^{f\leftarrow i}(t_{f},t_{i})\nonumber\\
&=&\delta_{fi}  -\frac{i}{\hbar} [\int dt_{1} <\phi^{a}_{f}(t_{1})|V_{vel.}(t_{1})|\phi_{i}^{a}(t_{1})>\nonumber\\
&+& \int \int dt_{1} <\phi^{a}_{f}(t_{1})|V_{vel.}(t_{1}) G_{a}(t_{1},t_{2})|\phi_{i}^{a}(t_{2})> \nonumber\\ 
&+& \cdots];  \lambda=0\label{TraAmpLamZer} \\
&=& \delta_{fi}  -\frac{i}{\hbar} [\int dt_{1} <\phi^{a}_{f}(t_{1})|V_{len.}(t_{1})|\phi_{i}^{a}(t_{1})>\nonumber\\
&+ &\int \int dt_{1}dt_{2} <\phi^{a}_{f}(t_{1})|V_{len.}(t_{1})
G_{a}(t_{1},t_{2})\nonumber\\
&\times&V_{len.}(t_{2})|\phi_{i}^{a}(t_{2})> + \cdots]; \lambda=1\label{TraAmpLamOne} \\
&=& \delta_{fi}  -\frac{i}{\hbar} [\int dt_{1} <\phi^{a}_{f}(t_{1})|V_{hyb.}(t_{1})|\phi_{i}^{a}(t_{1})>\nonumber\\
&+& \int \int dt_{1}dt_{2} <\phi^{a}_{f}(t_{1})|V_{hyb.}(t_{1})G_{a}(t_{1},t_{2})\nonumber\\
&\times&V_{hyb.}(t_{2})|\phi_{i}^{a}(t_{2})> + \cdots];  \lambda=\frac{1}{2}\label{TraAmpLamHal} 
\end{eqnarray}
where,  from Eq. (\ref{ParIntHam}),
\begin{eqnarray}\label{ThrInt}
V_{vel.}(t) &=& (-\frac{e}{mc}\bm{A}(t)\cdot\bm{p}_{op} + \frac{e^{2}}{2mc^{2}}A^{2}(t)); \lambda=0\nonumber\\
V_{len.}(t) &=& (\frac{e}{c}\dot{\bm{A}}(t)\cdot\bm{r})= (-e\bm{E}(t)\cdot\bm{r}); \lambda=1\nonumber\\
V_{hyb.}(t) &=&(\frac{1}{2}[(V_{vel.}(t) +V_{len.}(t)) -\frac{e^{2}}{4mc^{2}}A^{2}(t)]); \lambda=\frac{1}{2}\nonumber\\
\end{eqnarray}
The first two series (\ref{TraAmpLamZer}) and (\ref{TraAmpLamOne})
are readily recognized, on comparing  the expressions of the respective interactions in Eqs. (\ref{ThrInt}), as the two well-known
perturbation series, traditionally obtained in the ``velocity'' and the ``length'' gauge, respectively. 
Thus, they are seen to be nothing but 
two equivalent infinite series representations (branches) of the {\it same} amplitude function $S_{MC}^{f\leftarrow i}(t_{f},t_{i})$ and, hence, are but equal in their {\it common} domain of convergence (assuming, it exists) \cite{foot-02}. We may add that at present, little, if any thing, is known regarding the convergence properties of these infinite series representations of the same amplitude function. The third expansion (\ref{TraAmpLamHal}) provides another equivalent series representation (in terms of a ``hybrid interaction'') of the same amplitude by choosing $\lambda=\frac{1}{2}$, and so on for any other equivalent series.
Before concluding, we point out  that there is no difficulty in deriving a ``master'' expansion, analogous to Eq. (\ref{TraAmp}), from
the total Hamiltonian given in the ``length'' gauge (or for that matter in any other gauge) by proceeding exactly analogously as shown above -- it leads to the same conclusion of the mathematical equivalence of the two traditional perturbation series, in the velocity and length gauges, in a common domain of convergence, as demonstrated above.

To conclude, starting explicitly in the minimal-coupling (MC) ``velocity'' gauge,  we have derived a ``master'' perturbation expansion (\ref{TraAmp}) for generating a one-parameter family of equivalent infinite series representations of a given transition amplitude. The result is used to demonstrate (Eqs. (\ref{TraAmpLamZer}) and (\ref{TraAmpLamOne}))
that the two well-known perturbation series, traditionally obtained in the ``velocity'' and ``length'' gauges, are only two equivalent infinite series representations or branches of the same amplitude function $S_{MC}^{f\leftarrow i}(t_{f},t_{i})$, and hence, are equal in their {\it common} domain of convergence (provided that the latter exists).


\begin{references}
\bibitem{Lamb}
W. E. Lamb Jr.,  Phys. Rev. $\bm{85}$, 259 (1952).

\bibitem{Yang}
K.-H. Yang, Ann. Phys. (N.Y.) $\bm{101}$, 62 (1976).

\bibitem{Cohen-Tannoudji}
C. Cohen-Tannuodji, B. Diu and F. Lalo\"e, {\it  Quantum Mechanics} (Hermann/Wiley, Paris, 1977). 

\bibitem{Fock}
V. Fock, Zeit. f\"ur Physik $\bm{39}$, 226 (1926).

\bibitem{Faisal}
F.H.M. Faisal , J.Phys. B $\bm{40}$, F145 (2007); Phys. Rev. A $\bm{75}$,063412 (2007).

\bibitem{foot-01}
In the present context, the general field-free condition is
$\bm{E}(t_{i,f}) = -\frac{1}{c}\dot{\bm{A}}(t_{i,f})=0$, 
for $\bm{A}(t_{i,f})\equiv$ {\it arbitrary constants} or $0$, for any given values of $t_{i,f}$ (including the asymptotic values $t_{i,f}=\mp\infty$).
The general field-free atomic reference states are given by the fundamental
solutions of the Schr\"odiger equation associated with the reference Hamiltonian (\ref{ParHam}):
$H_{\lambda}(t_{i,f})=(\bm{p}_{op} -\frac{e}{c}\lambda\bm{A}(t_{i,f}))^{2}/2m + V_{a}(\bm{r})$; the corresponding initial and final reference states are thus the same as given in Eqs. (\ref{IniSta}) and (\ref{FinSta}) in the text.

\bibitem{foot-02}
This may be illustrated by the following  two 
well-known infinite series representations of the same plane wave function:
\begin{eqnarray}
e^{ikr\cos{\theta}}
&=& \sum_{n=0}^{\infty}(ikr\cos{\theta})^{n}/n!\nonumber\\
&=&1+ (ikr\cos{\theta})/1! - (kr\cos{\theta})^{2}/2! +\cdots.\nonumber\\
&=&\sum_{n=0}^{\infty} (2n+1)i^{n} j_{n}(kr) P_{n}(\cos{\theta})\nonumber\\
&=& j_{0}(kr) + (3 i)j_{1}(kr)P_{1}(\cos{\theta})- 5 j_{2}(kr)P_{2}(\cos{\theta})\nonumber\\
&+&\cdots. \label{PlaWav} \end{eqnarray}

We note that the Taylor series 
representation in the first line of (\ref{PlaWav}) above (which converges for $|kr\cos\theta | < 1$)
and the Bessel-Legendre series representation in the third line, 
have in general different domains of convergence.
They constitute two series representations (branches) 
of the plane wave function appearing on the left hand side; they are only equal at any point within the 
common domain of convergence
(as can be checked for example at a point  satisfying e.g.
$|kr\cos\theta|<1, (\theta\ne 0,\pi)$).
It goes without saying that their partial sums up to the same number of terms (e.g. the second and the fourth lines) may not necessarily agree numerically.
Finally, It is worth noting that a given asymptotic series could be often usefully transformed into another infinite series having an extended domain and/or a faster rate of convergence, for example, by a Shank's transformation 
\cite{Bender}.

\bibitem{Bender}
C.M. Bender and S.A. Orgzag, {\it Advanced Mathematical Methods for Scientists and Engineers} (McGraw-Hill, New York, 1978).

\end{references}
\end{document}